\documentclass{article}

\usepackage{fullpage, fourier, charter} 
\usepackage{graphicx} 
\usepackage{amsthm,amssymb}
\usepackage{xfrac}
\usepackage{authblk}
\usepackage{verbatim}
\usepackage{mathtools}
\usepackage[colorlinks=true,citecolor=blue]{hyperref}

\numberwithin{equation}{section}

\usepackage{color}

\title{Data-driven Soliton Manifold Approximations for Dark and 
Bright Waves: Some Prototypical 1d Case Examples}

\author[1]{Su Yang\thanks{Email: suyang@umass.edu}}
\author[1]{Shaoxuan Chen}
\author[2]{Wei Zhu}
\author[1,3,4]{Panayotis G.~Kevrekidis}

\affil[1]{Department of Mathematics and Statistics, University of Massachusetts Amherst, Amherst, MA 01003-4515, USA}
\affil[2]{School of Mathematics, Georgia Institute of Technology, Atlanta, GA 30332, USA}
\affil[3]{Department of Physics, University of Massachusetts Amherst, Amherst, MA 01003-4515, USA}
\affil[4]{Theoretical Sciences Visiting Program, Okinawa Institute of Science and Technology Graduate University, Onna, 904-0495, Japan}

\date{\small \today}

\begin{document}

\maketitle

\begin{abstract}
    In this paper, we revisit the investigation of 
    solitary-wave interactions in the nonlinear Schr\"odinger model, both in the presence and absence of a parabolic trapping potential. While approximate dynamics, based on variational ---or similar---
    methods, governed by a system of ordinary differential equations (ODEs) for both bright and dark-soliton interactions have been well established in the literature
    based on physical expert considerations,
    this study focuses on a data-driven approach, the so-called Sparse Identification of Nonlinear Dynamics (SINDy).
    Accordingly, our purpose is to use PDE time-series of select waveform diagnostics in order to numerically reconstruct such approximate dynamics, without prior knowledge thereof. The purpose is not only to verify the robustness of the dynamical approximated ODEs, but also to shed light on the application of such a data-driven methodology in the study of soliton interactions
    and to formulate a complementary approach, more reliant on
    the wealth of PDE data and less so on expert theoretical
    constructs.   
\end{abstract}

\textbf{Keywords.} Solitons, Data-driven method, Nonlinear Schr\"odinger model, SINDy.

\tableofcontents

\section{Introduction}\label{sec: introduction}

Solitary waves \cite{Ablowitz_Clarkson_1991, doi:10.1137/1.9781611970883, PhysRevE.64.026601, doi:10.1137/1.9781611973945} constitute ubiquitous nonlinear structures in numerous mathematical physics and biology models,
when the interplay of dispersion and nonlinearity is at hand. Some well-known models such as the Korteweg–De Vries (KdV) equation and nonlinear Schr\"{o}dinger equation (NLS) serve as prototypical 
partial differential equations of the dispersive nonlinear type for modeling solitary-wave interactions. 
Indeed, their generic emergence from other models, 
through multiple scale
expansions~\cite{ZAKHAROV1986455}, renders them 
practically ``normal form'' type scenarios for 
associated dynamics.

In recent years, there has been a growing interest within the nonlinear-wave community towards studying the interactions between multiple solitons \cite{suret2023solitongastheorynumerics, Ma_2016,PhysRevE.91.032905}; indeed the relevant topic
has been also extensively addressed in both 
reviews~\cite{Malomed2002Variational}, as well
as books~\cite{CarreteroFrantzeskakisKevrekidis2024NonlinearWaves}. Notably, the emerging concept of \textit{soliton gas} has been rigorously defined in the context of integrable systems as the asymptotic limit of the $N$-soliton solution \cite{suret2023solitongastheorynumerics}. It has been demonstrated that soliton interactions within a soliton gas can be effectively studied through both numerical and quasi-analytical approaches \cite{El_2021, PhysRevE.103.042201, FLAMARION2024114495, Bonnemain_2025, Bonnemain_2022}. 
In such settings, a large cohort of solitary waves
dynamically evolves through a dense array of interaction
events.

A complementary, time-honored approach over many
decades~\cite{MANTON1979397,kivshar,KIVSHAR1995353,higherd,Malomed2002Variational,Frantzeskakis_2010,Ma_2016} has been
to try to characterize the dynamics of few (typically two,
and building from there) solitary waves, extending then
considerations to larger clusters of structures; see, 
e.g.,~\cite{PhysRevE.91.032905}. It is important to 
appreciate that such low-dimensional descriptions have
been of particular value not only in the context of
one-dimensional structures such as solitary waves,
but also in higher-dimensional settings, as in the 
realm of vortices~\cite{doi:10.1137/07068597X,aref,Xie2018MultiVortexCrystal}, or vortex rings~\cite{konstantinov,caplan}.
Accordingly, direct numerical simulations have been
extensively complemented by (and compared to) reduced-order modeling of soliton (as well as vortex etc.) interactions. 
Such a reduction of the original partial differential
equation (PDE)
dynamics to the (here referred to as) ``soliton manifold''
plays an important role in understanding the underlying dynamics and has been used for building ``super-structures'',
including hypersolitons [solitons on lattices built out of
soliton interactions]~\cite{Ma_2016}, vortex
crystals [stationary patterns of many vortices~\cite{doi:10.1137/07068597X,aref}] etc.  This approach involves approximating the evolution of key soliton characteristics via a system of ordinary differential equations (ODEs), a feature that may not just be
theoretically attainable, but also 
experimentally probed, e.g., for bright~\cite{Nguyen2014CollisionsMatterWaveSolitons},
as well as for dark~\cite{Theocharis} solitary structures.

So far, 
for the nontrivial task of deriving such 
 effective ``particle like''
descriptions, we have been heavily relying on physical
expert approaches. These involve a variety of techniques,
including notably ones that are based on the variational structure
of the problem and the selection of a suitable
ansatz~\cite{Malomed2002Variational} (see also~\cite{CarreteroFrantzeskakisKevrekidis2024NonlinearWaves}
for a recent discussion). Here the ansatz incorporates
some of the key features of the wave, such as its 
amplitude, width, center position and velocity, among
others, and a substitution of the ansatz in the model's
Lagrangian or Hamiltonian enables the usage of 
Euler-Lagrange or Hamilton type equations for the 
effective time-dependent parameters. This represents effectively a ``projection'' (rigidly selected by an
expert) of the partial differential equation (PDE)
dynamics on the soliton manifold. While this appears
restricted to Hamiltonian systems, extensions do 
systematically exist for open dynamical settings, such
as, e.g., through the work of~\cite{galley}.
On the other hand, there exist similar techniques that
are based on asymptotic projection methods at the level
of the original partial differential
equations~\cite{E1994VorticesGL,Xie2018MultiVortexCrystal}
(e.g., in the context of vortices) that also
yield similar results. Similar results are also
obtained based on soliton perturbation theory~\cite{kivshar}.
In all of these classes of methods, expert ansatz
selection and detailed mathematical (variational, asymptotic or perturbative) analysis is 
leveraged to obtain the effective ODEs for the
wave interactions.

Here, we would like to advocate an alternative,
complementary, data-driven perspective toward this
problem of obtaining a suitable dynamical projection to the
``soliton manifold'' and the resulting ODEs.
Indeed, discovering governing equations has become a central theme in the field of dynamical systems. Various numerical algorithms and machine learning methods \cite{Brunton_2016, chen2019neuralordinarydifferentialequations, RAISSI2019686, doi:10.1137/1.9781611974508, Williams_2015} have been proposed for this purpose and all of them have proven to be 
valuable, under suitable circumstances. In this work, we focus on the methodology referred to as Sparse Identification of Nonlinear Dynamics (SINDy)~\cite{Brunton_2016}, a sparse regression-based algorithmic framework designed to identify parsimonious dynamical models from time series data. Specifically, as a prototypical benchmark problem of 
its kind, we apply SINDy to discover reduced ODE models governing soliton interactions in the nonlinear 
Schr{\"o}dinger system. In addition to the standard SINDy formulation, we also explore the use of its extension, SINDy with control \cite{fasel2021sindycontroltutorial}, which incorporates external inputs or known constraints during the identification process. Our intention is to illustrate
both the potential successes, but also the possible
problems that such a mainstream data-driven 
algorithmic platform may encounter in performing fundamental
soliton manifold projections/deriving solitonic ODE 
dynamics. In so doing, we hope to elucidate both some
of the technical steps needed to bring the methodology to the realm
of applicability, the comparison with some well-benchmarked
theoretical results and to highlight some of the improvements
that will be needed in order to further such a data-driven
methodology to the realm of problems that are considerably
more complex/not easily accessible theoretically.

The outline of this paper is arranged as follows. In section \ref{sec: model describe}, we provide a detailed description of the model which is considered in this work. In addition, we review some well-known analytical soliton solutions (both dark and bright solitons), and we also discuss some reduced dynamical systems which some particular features of the solitons satisfy. Finally, we also construct the multi-soliton initial data which shall be utilized in our numerical experiments. Next, in section \ref{sec: Data-driven method}, we briefly review (for completeness)
the  SINDy algorithm which is the central
methodological tool used for the data-driven derivation
of solitonic ODEs within this work. In section \ref{sec: numerical results}, we showcase the numerical results obtained in connection to the dynamical systems identification by applying SINDy. Here, we highlight both the cases where
SINDy works quite well (in comparison to the well-known
benchmark cases), but also the ones where it doesn't and
endeavor to rationalize such less successful outcomes
and when the interested practitioners may expect ones such.
Finally, 
Section~\ref{sec: Conclusions} concludes the paper with a summary of findings and suggestions for a number of
emerging research directions in this theme at the nexus
of data science and nonlinear waves that we believe merits
further study.

\section{Model description}\label{sec: model describe}

The model we will consider in this paper is the one-dimensional Gross-Pitaevskii (GP) equation, which serves as a variant of the standard NLS equation with an external potential. 
This is the canonical mean-field model for 
ultracold atoms in the presence of external 
confinement~\cite{PitaevskiiStringari2016,kevrekidis2015defocusing}, although it is also of relevance in nonlinear
optics in the presence of spatially varying 
refractive index~\cite{Malomed2002Variational}.
Specifically,  we study the following equation:

\begin{equation}\label{eq: nonlinear schrodinger model}
    iu_t = -\frac{1}{2}u_{xx} + g\left|u\right|^{2}u + V_{\text{MT}}(x)u,
\end{equation}
where $g = \pm 1$ determines the focusing ($g = -1$) or defocusing ($g = +1$) nonlinearity, and $V_{\text{MT}}(x) = \frac{1}{2}\Omega^{2}x^{2}$ refers to a harmonic (magnetic) trapping potential with a real constant $\Omega$. Unless otherwise specified, we fix $\Omega = 0.025$ [motivated
by the feature that $\Omega$ represents a ratio of
longitudinal to transverse confinement and for the 
relevant reduction to be valid needs to be $\Omega \ll 1$]
in all numerical experiments presented in this paper.

\subsection{Dark soliton}

When $g = 1$ and in the absence of the trap, i.e., for 
$V_{\text{MT}}=0$, it is well-known~\cite{Frantzeskakis_2010} that the model \eqref{eq: nonlinear schrodinger model} admits the following exact dark soliton (DS) solution:
\begin{equation}\label{eq: dark soliton solution}
    u_{\text{ds}}\left(x,t\right) = \sqrt{n_0}\left[B\text{tanh}\{\sqrt{n_0}B\left[x - \xi\left(t\right)\right]\} + iA\right]e^{i\phi_{\text{ds}}\left(t\right)},
\end{equation}
where $n_0 \in \mathbb{R}^{+}$ is the density of the constant background supporting the dark soliton, and $\xi\left(t\right)=vt+\xi_0$ denotes the soliton position (with $\xi_0$ the initial location and $v$ the velocity). The associated phase is
\begin{equation}\label{eq: phase def}
    \phi_{\text{ds}}\left(t\right) = -n_{0}t + \phi_0,
\end{equation}
with $\phi_0$ being the initial phase. The parameters $A$ and $B$, together with the soliton velocity $v$, satisfy the relations $A^{2} + B^{2} = 1$ and $v = A\sqrt{n_0}$.

Assume there exist $N_s \in \mathbb{Z}^{+}$  dark solitons in a multi-dark-soliton state, then it is known that the dynamics of their positions approximately obey the following second-order ODE~\cite{KIVSHAR1995353,Frantzeskakis_2010,Ma_2016}:
\begin{equation}\label{eq: dark solitary wave dynamics ODE}
    \ddot \xi_{i} = 8n_0^{3/2}e^{2\sqrt{n_0}\left(\xi_{i-1} - \xi_{i}\right)} - 8n_0^{3/2}e^{2\sqrt{n_0}\left(\xi_i - \xi_{i+1}\right)}-\omega^2\xi_i,
\end{equation}
where $\omega = \Omega/\sqrt{2}$, $\xi_{i}$ denotes the position of the $i$-th dark soliton in the chain, and $\ddot \xi_i= \frac{d^{2}}{dt^{2}}\xi_{i}$.

\subsection{Bright soliton}\label{subsec: BSs description}

On the other hand, when $g = -1$ and in the absence of the external potential (i.e., $V_{\text{MT}} = 0$), the one-dimensional GP equation \eqref{eq: nonlinear schrodinger model} admits the following bright soliton (BS) solution \cite{Ma_2016},
\begin{equation}\label{eq: BS solution}
    u_{\text{bs}}(x,t) = a\hspace{0.5mm}\text{sech}\bigg[a(x-\xi(t))\bigg]\exp\bigg[i\left(vx+\phi_{\text{bs}}(t)\right)\bigg],
\end{equation}
where $a$ is the amplitude of the bright soliton, $\xi(t) = \xi_0 + vt$ denotes the position, and $\phi_{\text{bs}} = \phi_0 + \frac{\left(a^{2}-v^{2}\right)t}{2}$ is the associated phase. Here, $\xi_0$ and $\phi_0$ represent the initial position and phase  of the bright soliton, respectively.

Moreover, consider a chain of  bright solitons, where the parameters of the $i$-th soliton are described by the vector $P_i = (a_i,\xi_i,v_i,\phi_i)$, where the four parameters represent the amplitude, position, velocity, and phase of the $i$-th soliton, respectively. It has been shown that the dynamical evolution of these parameters ought to be approximately described by the following system of first-order ODEs:
\begin{equation}\label{eq: BS dynamical ODEs BSs}
   \begin{aligned}
    &\dot a_j = 4a_j^{2}\left(S_{j,j-1} - S_{j,j+1}\right),\\
    &\dot v_j = -4a_j^{2}\left(C_{j,j-1}-C_{j,j+1}\right),\\
    &\dot\xi_j = v_j - 2\left(S_{j,j-1}+S_{j,j+1}\right),\\
    &\dot\phi_j = \frac{a_j^{2}+v_j^{2}}{2} - 2v_j\left(S_{j,j-1}+S_{j,j+1}\right) + 6a_j\left(C_{j,j-1}+C_{j,j+1}\right),
    \end{aligned}
\end{equation}
where 
\begin{equation}
    \begin{aligned}
       &S_{j,n} = \exp\bigg[-\bigg|a_n(\xi_j-\xi_n)\bigg|\bigg]a_n\sin(s_{j,n}\phi_{j,n}),\\
       &C_{j,n} = \exp\bigg[-\bigg|a_n(\xi_{j}-\xi_{n})\bigg|\bigg]a_n\cos(\phi_{j,n}),\\
       &\phi_{j,n} = \phi_j - \phi_n - v_n(\xi_j-\xi_n),\\
       &s_{j,j-1} = -s_{j,j+1} = 1;
    \end{aligned}
\end{equation}
see for further details~\cite{Ma_2016}, as well
as references therein.
However, it is important to note that the system \eqref{eq: BS dynamical ODEs BSs} is valid only under the following conditions: (i) all bright solitons considered need to have similar amplitudes and velocities. Mathematically, this requires that $|a_i - a_j| \ll \bar{a}$ and $|v_i-v_j| \ll \bar{v}$, where $\bar{a},\bar{v}$ denote the average amplitude and velocity of the bright solitons. (ii) All adjacent bright solitons need to be sufficiently well separated; this means that $\bar{a}|\xi_i-\xi_j| \gg 1$. (iii) All bright solitons need to satisfy the height-separation constraint: $|a_i-a_j||\xi_i-\xi_j| \ll 1$. 

Moreover, when the three conditions (i)--(iii) are fulfilled, the system of ODEs in \eqref{eq: BS dynamical ODEs BSs} can be, to a further degree of approximation (essentially
rendering all $a_i$'s equal), reduced to the following second-order ODE,
\begin{equation}\label{eq: BS reduced dynamical ODE}
    \ddot\xi_{i} = \sigma_{i-1,i}4a^{3}\exp\left(-a(\xi_i-\xi_{i-1})\right) - \sigma_{i,i+1}4a^{3}\exp\left(-a(\xi_{i+1}-\xi_{i})\right),
\end{equation}
where $\sigma_{i,j} = \pm1$ is determined by the relative phase of the two consecutive bright solitons. Specifically, when the solitons are out of phase (OOP), i.e., $|\phi_i - \phi_j| = \pi$, we have $\sigma_{i,j} = 1$, while $\sigma_{i,j} = -1$ when the solitons are in phase (IP), i.e., $\phi_i = \phi_j$. Moreover, for OOP bright solitons, when there exists a trap (i.e. $\Omega \neq 0$), the reduced dynamics of soliton interaction reads,
\begin{equation}\label{eq: OOP bright solitons interaction with trap}
    \ddot\xi_i = 4a^3\exp\left(a\left(\xi_{i-1}-\xi_i\right)\right) - 4a^3\exp\left(a\left(\xi_i-\xi_{i+1}\right)\right) - \Omega^2\xi_i;
\end{equation}
i.e., the trap adds a longitudinal harmonic confinement
of frequency $\Omega$ to each of the solitary waves~\cite{Ma_2016}.

\subsection{Multi-soliton construction}

To study the interaction dynamics of the bright or dark solitons in the GP equation \eqref{eq: nonlinear schrodinger model}, it is essential to construct appropriate initial data which we call the multi-soliton initial conditions. The procedures for generating these initial conditions differ between the bright and dark soliton cases.

For a chain of $N$ dark solitons, the initial condition 
for the i-th dark soliton $u_{\text{ds}}^{(i)}$ leverages Eq.~\eqref{eq: dark soliton solution}. Specifically, we define the initial state of $N$ dark solitons through a product
ansatz as:
\begin{equation}\label{eq: dark-soliton chain}
    u_{\text{ds}}(x,0) = \prod_{i = 1}^{N}u_{\text{ds}}^{(i)}(x,0).
\end{equation}

On the other hand, for the bright soliton case, we focus exclusively on OOP bright solitons, where the phases difference between adjacent solitons is $\pi$. In this scenario, the initial state of the bright soliton chain is constructed by summing the individual bright soliton profiles defined in Eq.~\eqref{eq: BS solution}. Namely, in this work, we utilize the following initial state for the bright soliton chain
\begin{equation}\label{eq: BS chain}
    u_{\text{bs}}\left(x,0\right) = \sum_{i = 1}^{N}\left(-1\right)^{i-1}\text{sech}\left(x - \xi_i\right),
\end{equation}
where each soliton is assumed to have constant unit amplitudes with zero velocities so as to fulfill the conditions (I), (II), and (III) described in sub-section \ref{subsec: BSs description}.

\section{Data-driven methods}\label{sec: Data-driven method}

In this section, we briefly review the data-driven method known as Sparse Identification of Nonlinear Dynamics (SINDy)~\cite{Brunton_2016}, which serves as the primary framework for identifying soliton-interaction dynamics in the
present setting. In addition to the standard SINDy approach, we also discuss an extended variant, SINDy with control~\cite{fasel2021sindycontroltutorial}, which is likewise a relevant methodology for our study. {Essentially, what this method achieves is that users have the flexibility to choose their customized time-series dataset (augmenting
the ones of the state variables) which can be directly added to the library, and these customized time-series data are the so-called control inputs.}

\subsection{Sparse identification of nonlinear dynamics}\label{subsec: SINDy}

SINDy \cite{Brunton_2016} is a powerful data-driven method 
whose scope is to discover governing ODEs (or PDEs, where
appropriate) based on simulated or observational data and the
selection of suitable user-selected/supplied libraries. Mathematically, we consider the following dynamical system
\begin{equation}\label{eq: dynamical ODE}
    \frac{dx}{dt} = f(x),
\end{equation}
where $x(t) = (x_1(t),x_2(t),\ldots,x_n(t)) \in \mathbb{R}^{n}$ is the state variable at the particular time $t$, and $f:\mathbb{R}^{n} \to \mathbb{R}^{n}$ refers to a nonlinear mapping which governs the evolution dynamics of the state $x$. The goal of SINDy is to identify an approximate representation of the (unknown) governing function $f$ given a time series of state observations $x(t)$. 

The fundamental assumption of SINDy is that the nonlinear mapping $f$ has a ``simple'' expression, allowing it to be approximated as a linear combination of a small number of terms from a prescribed library, denoted as $\theta(x) = \left[\theta_1(x), \theta_2(x), \ldots, \theta_p(x)\right]$, where each $\theta_j(x),  j \in [1,2,\ldots,n]$ is a nonlinear candidate feature function. Typical choices for $\theta_j(x)$ include sinusoidal, polynomials, exponentials, and other nonlinearities commonly encountered in physical models.
The user can also supply relevant terms, based on 
physical expertise when that is available for the system
in question.

In particular, SINDy assumes the right-hand side nonlinear dynamics of Eq.~\eqref{eq: dynamical ODE} can be approximated in the following way
\begin{equation}\label{eq: SINDy approximation}
    f(x) = \bigg[f_1(x), f_2(x), \ldots, f_n(x)\bigg] \approx \theta(x) \cdot \Xi = [\theta_1(x),\theta_2(x),\ldots,\theta_p(x)] \cdot \bigg[\xi_1,\xi_2,\ldots,\xi_n\bigg],
\end{equation}
where $\Xi = \bigg[\xi_1,\xi_2,\ldots,\xi_n\bigg] \in \mathbb{R}^{p\times n}$ is a sparse matrix. Each column $\xi_i \in \mathbb{R}^p$ determines which terms from the library $\theta(x)$ are selected to represent the component function $f_i(x)$ of the dynamics. 

To identify the sparse coefficient matrix $\Xi$, SINDy performs sparse regression on the data. Specifically, we consider the time-series data of the state $x(t)$, given by $\left[x(t_1), x(t_2), \ldots, x(t_N)\right] \in \mathbb{R}^{n\times N}$ at discrete time snapshots of $t_1, t_2, \ldots, t_N$. Then we have the following data matrix and corresponding time-derivative matrix:
\begin{equation}\label{eq: Data and derivative matrices}
    \begin{aligned}
        &X = \left[x(t_1), x(t_2), \ldots, x(t_N)\right]^{T} = \bigg[X_1,X_2,\ldots,X_n\bigg] = \begin{bmatrix}
            x_1(t_1) & x_2(t_1) & \ldots & x_n(t_1)\\
            x_1(t_2) & x_2(t_2) & \ldots & x_n(t_2)\\
            \vdots & \vdots & \ddots & \vdots \\
            x_1(t_N) & x_2(t_N) & \ldots & x_n(t_N)
        \end{bmatrix} \in \mathbb{R}^{N\times n}, \\
        &\frac{dX}{dt} = \bigg[\dot X_1, \dot X_2, \ldots, \dot X_n\bigg] = \begin{bmatrix}
            \dot x_1(t_1) & \dot x_2(t_1) & \ldots & \dot x_n(t_1)\\
            \dot x_1(t_2) & \dot x_2(t_2) & \ldots & \dot x_n(t_2)\\
            \vdots & \vdots & \ddots & \vdots \\
            \dot x_1(t_N) & \dot x_2(t_N) & \ldots & \dot x_n(t_N)
        \end{bmatrix} \in \mathbb{R}^{N\times n},
    \end{aligned}
\end{equation}
where the derivative matrix $\frac{dX}{dt}$ can, for example, be numerically approximated from the data matrix $X$ using a finite-difference scheme [naturally, any derivative 
approximation method will involve a certain error, with
the numerical method aiming to minimize such error]. We then define the library matrix $\Theta(X) \in \mathbb{R}^{N \times p}$ as follows
\begin{equation}\label{eq: Library matrix}
    \Theta(X) = \bigg[\theta_1(X), \theta_2(X), \ldots, \theta_p(X)\bigg].
\end{equation}
Combining Eq.~\eqref{eq: dynamical ODE} and \eqref{eq: SINDy approximation} yields the following problem
\begin{equation}\label{eq: Optimization problem}
    \frac{dX}{dt} \approx \Theta(X) \Xi,
\end{equation}
To compute the sparse matrix $\Xi$, one may employ various techniques such as LASSO-type sparse regression, sequentially thresholded least-square regression \cite{doi:10.1073/pnas.1517384113} (STLSQ), or forward regression orthogonal least-squares \cite{article} (FROLS). 
In this work, we primarily adopt STLSQ and FROLS as the main optimization strategies for identifying the governing equations of soliton-interaction dynamics. 

It is worth noting that the optimizer FROLS is a greedy algorithm that incrementally selects terms based on their correlation with the residual. In all our numerical experiments involving FROLS, we limit the maximum number of iterations to two. This choice is motivated by the observation that, based on the approximated dynamics of the interactions of both bright and dark solitons, only a few terms are involved in the right-hand side of the ODEs. Therefore, we select only the most relevant terms that exhibit the strongest correlation with the input time series data, ensuring both model sparsity and interpretability.

\subsection{SINDy with controls}

SINDy with controls~\cite{fasel2021sindycontroltutorial} extends the standard SINDy framework described in Subsection~\ref{subsec: SINDy}. In particular, we consider the following nonlinear dynamical system:
\begin{equation}\label{eq: Nonlinear dynamics with control}
    \frac{dx}{dt} = f(x,u),
\end{equation}
where the $u$ variable denotes the input control.

Given the time-series data of both the state $x$ and the control $u$, Eq.~\eqref{eq: Optimization problem} becomes
\begin{equation}\label{eq: Optimization problem of SINDy with control}
    \frac{dX}{dt} \approx \Theta(X,U)\Xi, 
\end{equation}
where $U = [u_1,u_2,\ldots,u_n]^{T}$ is the data set of the controls. 

As in the standard case, performing sparse regression on Eq.~\eqref{eq: Optimization problem of SINDy with control} yields the solution for the sparse matrix $\Xi$ that describes the dynamics.

\section{Numerical results}\label{sec: numerical results}

In this section, we present numerical results on identifying soliton-interaction dynamics using both the standard SINDy method and its control-augmented variant. We divide this section into three subsections: Subsection~\ref{eq: subsec data generation} provides some preliminaries on how our dataset are processed and generated. Subsection~\ref{subsec: dark solitons results} focuses on the data-driven discovery of interaction dynamics for dark solitons, while Subsection~\ref{subsec: bright solitons results} presents the corresponding results for bright solitons.

\subsection{Data generation}\label{eq: subsec data generation}

Before we display our numerical results on the identification of soliton interaction dynamics, it is important to first understand how our dataset is generated. Firstly, we note that we apply the RK4 time integration scheme with a finite-difference discretization of space to numerically solve Eq.~\eqref{eq: nonlinear schrodinger model}. On the one hand, for the generation of the data for the positions of both bright and dark solitons, we first numerically pinpoint the $x$ coordinates of all the global peaks 
(maxima or minima, respectively) of the intensity of the wave function (i.e. $|u(x,t)|^2$) and we shall treat the values of these $x$ coordinates as the rough estimates of the locations of the solitons. Notice that we  repeat such a process for every time snapshot $t$ so as to obtain the time-series data that we denote as $\widetilde{\xi}(t)$. However, it is not appropriate to directly use $\widetilde{\xi}(t)$ since it is only a rough estimate of the soliton positions and the smoothness of such a dataset is suboptimal. Based on $\widetilde{\xi}(t)$, we then perform a least-square fitting on the profile of the solitons, centered at $\widetilde{\xi}(t)$, to further enhance the smoothness of the time-series data of soliton positions. Finally, we obtain our dataset for the position of the solitons, which is denoted as $\xi(t)$ in this paper. On the other hand, the data for the velocity $v(t)$ of the solitons are simply obtained by performing finite difference on the position time-series data of $\xi(t)$. Thirdly, the data for amplitudes, denoted as $a(t)$, of the (bright) solitons are obtained by numerically computing the maximum of the absolute value of the wave function (i.e. $|u(x,t)|$). Finally, for the phase time-series data $\phi(t)$ of the solitons, we calculate and treat the phase angles of the wave function $u(x,t)$ centered at $\xi(t)$ as our data for $\phi(t)$.

\subsection{Interaction of dark solitons}\label{subsec: dark solitons results}

We begin by investigating the interaction dynamics of dark solitons using the standard SINDy framework. In particular, we examine cases involving two and four interacting dark solitons 
[such cases with even soliton numbers 
are conducive to the use of periodic boundary
conditions].
Here, we aim to identify the governing equations for their respective positions. Beyond varying the number of solitons in the chain, we also examine the effect of an external parabolic trap potential, with the goal of capturing how this additional influence modifies the underlying dynamics and contributes to the identification of accurate reduced-order models for dark soliton interactions.

\subsubsection{Two dark solitons without parabolic trap potential}\label{subsec: Two DSs without trap}

We start our investigation by examining the situation where only two dark solitons interact. For simplicity, we fix the background density to  $n_0 = 1$, although our results can
straightforwardly be extended to other values of $n_0$. The dynamics governed by Eq.~\eqref{eq: nonlinear schrodinger model} are simulated using a fourth-order Runge–Kutta method for time integration and a finite-difference scheme for spatial discretization, under periodic boundary conditions. This simulation generates the time-series data required as input for the SINDy algorithm. Also note that SINDy is only applicable for a first-order ODE system, so we rewrite Eq.~\eqref{eq: dark solitary wave dynamics ODE} as
\begin{equation}\label{eq: change into first-order system}
    \begin{aligned}
        \dot\xi_{i} &= v_{i},\\
        \dot v_{i} &= 8n_0^{3/2}e^{2\sqrt{n_0}\left(\xi_{i-1} - \xi_{i}\right)} - 8n_0^{3/2}e^{2\sqrt{n_0}\left(\xi_i - \xi_{i+1}\right)}.
    \end{aligned}
\end{equation}
where the term $-\omega^2\xi_i$ is missing due to the absence of the parabolic trap potential. Then, our input time-series data, denoted by $\mathbf{X}$, are
\begin{equation}\label{eq: Input time series data}
    \mathbf{X} = \left[\xi_1,\xi_2,\dot \xi_1, \dot \xi_2\right] = \left[\xi_1, \xi_2, v_1, v_2\right],
\end{equation}
where the two first-order derivatives of $\dot \xi_1, \dot \xi_2$ are computed numerically by the central difference scheme.

We numerically solve the Eq.~\eqref{eq: nonlinear schrodinger model} with the initial condition 
\eqref{eq: dark solitary wave dynamics ODE} but with three distinct sets of initial positions and velocities. The resulting time-series data, denoted by $\mathbf{X_1}, \mathbf{X_2}, \mathbf{X_3}$, are then concatenated vertically as:
\begin{equation}\label{eq: Input time-series}
    \mathbf{X} = 
    \begin{bmatrix}
        \mathbf{X}_1 \\
        \mathbf{X}_2 \\
        \mathbf{X}_3
    \end{bmatrix}.
\end{equation}
Motivated by the structure of the approximate dynamical equations in Eq.~\eqref{eq: change into first-order system}, we define the SINDy library to include the following three candidate terms: $\Theta = [x, x^2, \exp\left(2(x-y)\right)]$, where $x, y$ refer to the terms in the set of the time-series data: $\left[\xi_1,\xi_2,v_1,v_2\right]$. As a result, there are totally $8 + \binom{4}{2} = 14$ terms in the library. 
With the input time-series data \eqref{eq: Input time-series}, the SINDy prediction reads,
\begin{equation}\label{eq: SINDy prediction for 2 ds}
   \begin{aligned}
    \dot\xi_1 &= 1.000v_1,\\
    \dot\xi_2 &= 1.000v_2,\\
    \dot v_1 &= -7.990\exp\left(2\left(\xi_1 - \xi_2\right)\right),\\
    \dot v_2 &= 7.988\exp\left(2\left(\xi_1 - \xi_2\right)\right).
    \end{aligned}
\end{equation}
In this scenario, we notice that the SINDy prediction \eqref{eq: SINDy prediction for 2 ds} accurately recovers the approximate dynamics of the dark-soliton interactions described in Eq.~\eqref{eq: change into first-order system}. This is further illustrated in the left panel of Figure~\ref{fig:dark soliton positions comparison}, where the evolution of the two dark solitons predicted by SINDy closely aligns with the analytically derived system~\eqref{eq: dark solitary wave dynamics ODE}.
Crucially, such a result requires the use of concatenated time-series data generated from multiple distinct initial conditions. When SINDy is applied using data from only a single initial condition, the resulting model fails to capture the correct dynamics due to overfitting and deviates substantially from the accurate prediction in Eq.~\eqref{eq: SINDy prediction for 2 ds}.

\subsubsection{Four dark solitons without parabolic trap potential (no interaction across boundary)}

We now turn to the SINDy prediction for the interaction dynamics of four dark solitons. In this case, we do not account for the interaction between the first and last soliton, as the total simulation time is chosen to ensure that these solitons remain well separated from the domain boundaries and thus interact only negligibly. Indeed,
in the current setting
the outer soliton interaction is screened by the 
presence of the intermediate solitons and hence is
exponentially weaker than the dominant terms of interest.

\begin{figure}[t!]
    \centering
    \includegraphics[width=1\linewidth]{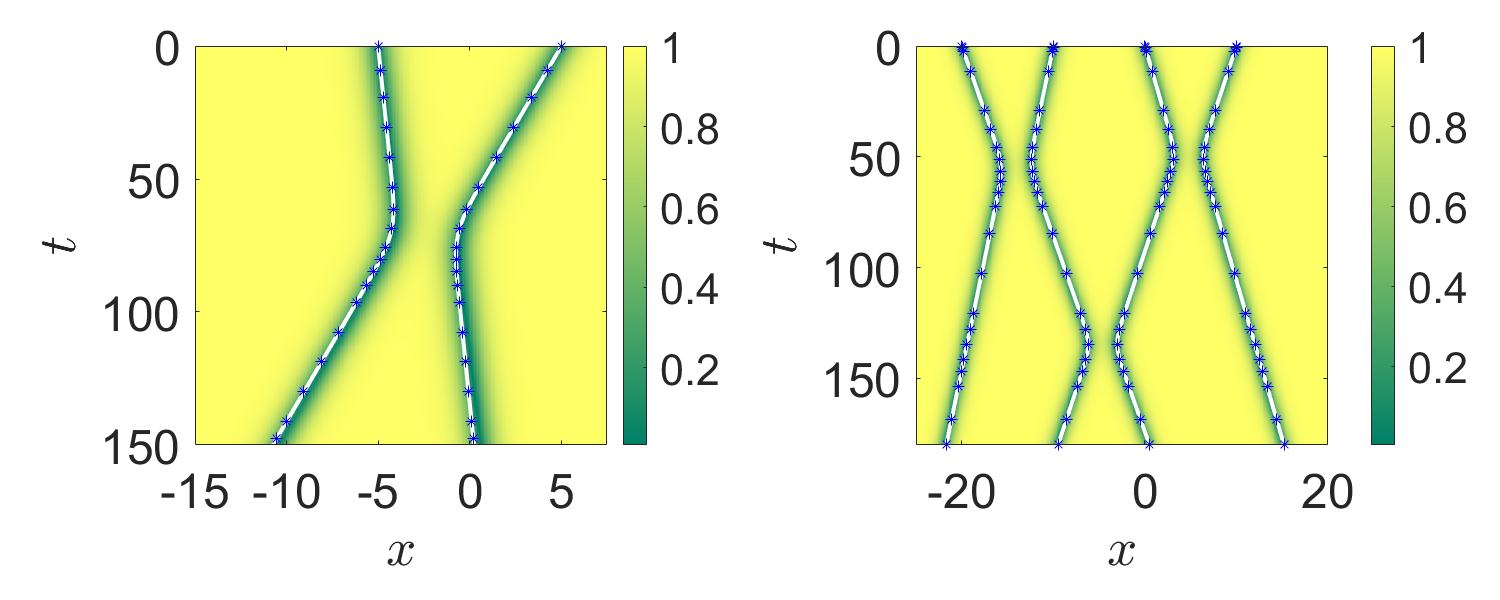}
    \caption{The spatio-temporal plot of dark-soliton interaction dynamics without a parabolic trap potential: the left panel shows the interaction of two dark solitons, while the right panel presents that of four dark solitons. Notice that the white solid curves and blue asterisks refer to theoretical prediction of the soliton positions based on system~\eqref{eq: dark solitary wave dynamics ODE} and on the SINDy prediction based on~\eqref{eq: SINDy prediction for 2 ds} (two dark solitons) and \eqref{eq: dynamics of 4 ds} (four dark solitons), respectively.}
    \label{fig:dark soliton positions comparison}
\end{figure}

Before applying SINDy for dynamical prediction, we sample time-series data from two distinct initial conditions and concatenate them in a manner similar to Subsection~\ref{subsec: Two DSs without trap}. Here, we notice that the library includes: $\Theta = \{x,\exp\left(2(x-y)\right)\}$, where $x,y \in \left[\xi_1,\xi_2,\xi_3,\xi_4,v_1,v_2,v_3,v_4\right]$, so 
in total there are $8 + \binom{8}{2} = 36$ terms.
The resulting SINDy prediction is then given by:
\begin{equation}\label{eq: dynamics of 4 ds}
    \begin{aligned}
        \dot \xi_1 &= 1.000 v_1,\\
        \dot \xi_2 &= 1.000 v_2,\\
        \dot \xi_3 &= 1.000 v_3,\\
        \dot \xi_4 &= 1.000 v_4,\\
        \dot v_1 &= -8.016\exp\left(2\left(\xi_1 - \xi_2 \right)\right),\\
        \dot v_2 &= 8.021\exp\left(2\left(\xi_1 - \xi_2 \right)\right) - 8.010\exp\left(2\left(\xi_2 - \xi_3\right)\right),\\
        \dot v_3 &= 8.011\exp\left(2\left(\xi_2 - \xi_3\right)\right) - 8.007\exp\left(2\left(\xi_3 - \xi_4\right)\right),\\
        \dot v_4 &= 8.018\exp\left(2\left(\xi_3 - \xi_4\right)\right).
    \end{aligned}
\end{equation}
Similarly, SINDy successfully recovers the analytically approximated dark-soliton interaction dynamics described in system \eqref{eq: dark solitary wave dynamics ODE}, with the right panel of Figure~\ref{fig:dark soliton positions comparison} providing further evidence for the close agreement between the analytically derived trajectories and those predicted by SINDy. It is perhaps worthwhile to note
in passing that both in this example and the previous one,
while the results were highly accurate as concerns
the prediction of interactions and well in line with
theoretical expectations from earlier works such as
as~\cite{KIVSHAR1995353,Theocharis,Coles_2010}, they 
did seem to involve slightly unequal coefficients,
which is slightly ``misaligned'' with Newton's third
(equal action-reaction) law. We will return to address
that point in subsequent sections.

\subsubsection{Four dark solitons without parabolic trap potential (with interaction across boundary)}

To explore the influence of boundary-induced interactions, we consider a longer simulation time that allows the first and last solitons to reach opposite boundaries and interact. To be more specific, we simulate the GP equation \eqref{eq: nonlinear schrodinger model} with a sufficiently large total simulation time so that both of these two dark solitons  can travel to the boundaries of the computational domain so that they can interact. The corresponding SINDy-identified model is:
\begin{equation}\label{eq: SINDy prediction on 4 DSs with 1 interacting with 4}
    \begin{aligned}
       \ddot \xi_1 &= -7.994\exp\left(2\left(\xi_1-\xi_2\right)\right) + 8.007\exp\left(-2\left|2L-\left|\xi_1-\xi_4\right|\right|\right),\\
       \ddot\xi_2 &= 8.002\exp\left(2\left(\xi_1-\xi_2\right)\right) - 7.994\exp\left(2\left(\xi_2 -\xi_3\right)\right),\\
       \ddot\xi_3 &= 8.001\exp\left(2\left(\xi_2-\xi_3\right)\right) - 8.001\exp\left(2\left(\xi_3 -\xi_4\right)\right),\\
       \ddot\xi_4 &= 8.005\exp\left(2\left(\xi_3-\xi_4\right)\right) - 8.001\exp\left(-2\left|2L-\left|\xi_1-\xi_4\right|\right|\right),
    \end{aligned}
\end{equation}
where $L = 30$ denotes one half of the length of the computational domain. We notice that the second term on the right-hand side of $\ddot\xi_1$ and $\ddot\xi_4$ in system \eqref{eq: SINDy prediction on 4 DSs with 1 interacting with 4} represents the interaction between the first and the last dark soliton. Notice that the library used in this example is: $\Theta = \{x,x^2,\exp\left(2(x-y)\right),\exp\left(-2|2L-|x-y||\right)\}$, 
i.e., it also involves the interaction term induced
due to the presence of the ``periodic boundary'', that
is it takes into account the fact that the simulation
occurs on a ring.
Notice that the SINDy with control method is applied to this example, where the control inputs include $\xi_1,\xi_2,\xi_3,\xi_4$, and the original input time-series are $\dot\xi_1,\dot\xi_2,\dot\xi_3,\dot\xi_4$ so this means there are totally $2\times8 + 2\times\binom{8}{2} = 72$ terms in the library.

\subsubsection{One dark soliton with parabolic trap potential}
\label{sec:one_ds_parabolic}

Next, we examine the case where a parabolic trap potential is incorporated into the GP equation~\eqref{eq: nonlinear schrodinger model}. In particular, we set the potential to $V_{\text{MT}} = \frac{1}{2}\Omega^2x^2$ with $\Omega = 0.025$.
We first investigate the position dynamics $\xi$ of a single dark soliton. %
The SINDy with control is applied in this example: the control variable is $\xi$, while the input data is $\dot\xi$. The library is $\Theta = \{\exp(2(x-y)), x, x^2\}$, so there are totally $2\times2+\binom{2}{2} = 5$ terms. Here, the
SINDy prediction yields:
\begin{equation}\label{eq: dynamics of a single DS}
    \ddot\xi = -0.00031287\xi.
\end{equation}
We first notice that the dynamics \eqref{eq: dynamics of a single DS} is learned by applying the SINDy with control via the FROLS optimizer, which is a greedy method. Motivated by the structure of the approximate dynamics in Eq.\eqref{eq: dark solitary wave dynamics ODE}, we focus on selecting a single dominant term on the right-hand side that governs the evolution of the dark soliton’s position. 

Furthermore, we construct a library:
{$\Theta = \{\exp(2(x-y)), x, x^2, \ldots, x^{10}\}$, where $x,y \in \{\dot\xi,\xi\}$. We notice that it is not physically meaningful to have interaction between the two terms of $\xi$ and $\dot\xi$ as these two quantities even have different units. Nevertheless, SINDy chooses to ignore relevant interactions between these two terms, and its prediction is still reasonable.} 
In summary, the SINDy model consistently and robustly recovers the expected 
single-soliton dynamics, accurately reflecting the analytical structure given of the reduced single-soliton ODE.

\subsubsection{Two dark solitons with parabolic trap potential}

In the process of sampling the time-series data used for SINDy implementation, we take $\Omega = 0.025$. We then apply the method of SINDy with control.
Here, the library is $\Theta = \{x,x^2,\exp\left(2(x-y)\right)\}$. The input time-series data are $[\dot\xi_1,\dot\xi_2]$, while the control variables are $[\xi_1,\xi_2]$. Hence, there are totally $2\times4+\binom{4}{2}=14$ terms.
The resulting dynamics, as predicted by SINDy reads:
\begin{equation}\label{eq: SINDy prediction for 2 ds with parabolic trap}
    \begin{aligned}
        \ddot \xi_1 &= -7.757\exp\left(2\left(\xi_1 - \xi_2\right)\right) - 0.00031635\xi_1,\\
        \ddot \xi_2 &= 7.803\exp\left(2\left(\xi_1 - \xi_2\right)\right) - 0.00031674\xi_2.
    \end{aligned} 
\end{equation}
It is evident that the SINDy prediction in Eq.\eqref{eq: SINDy prediction for 2 ds with parabolic trap} closely aligns with the approximated dynamics described by Eq.\eqref{eq: dark solitary wave dynamics ODE}.

Nevertheless, it should be highlighted that now
the deviations are more pronounced. This is due
to two different features. On the one hand,
the Newton's third law ``imbalance''
between the interaction terms is somewhat more
pronounced. On the other hand, here the interactions
are screened by the presence of the density modulation;
similar examples have appeared in the literature
also in the context of vortices when the density
is inhomogeneous, with a relevant example being,
e.g., the work of~\cite{busch}. This may be partially
reflected in the interaction coefficient deviation
from the homogeneous expectation of the prefactor
of $8$~\cite{KIVSHAR1995353}.

\subsubsection{Four dark solitons with parabolic trap potential}

We then test and extend the result of Eq.~\eqref{eq: SINDy prediction for 2 ds with parabolic trap} to the case of a chain where four dark solitons are involved. 
Here too, we utilize a library
involving $\Theta = \{x,x^2,\exp\left(2(x-y)\right)\}$. The input data are $\dot\xi_1,\dot\xi_2,\dot\xi_3,\dot\xi_4$, and the control variables are $\xi_1,\xi_2,\xi_3,\xi_4$, so there are totally $2\times 8 + \binom{8}{2} = 44$ terms.
Applying the method of SINDy with control again in this case, the predicted dynamics for the time-series $\left[\ddot\xi_1, \ddot\xi_2, \ddot\xi_3, \ddot\xi_4\right]$ reads, %
\begin{equation}\label{eq: dynamics for 4 ds with parabolic trap}
    \begin{aligned}
        \ddot \xi_1 &= -7.950\exp\left(2\left(\xi_1 -\xi_2\right)\right) - 0.00031272\xi_1,\\
        \ddot \xi_2 &= 8.420\exp\left(2\left(\xi_1 - \xi_2\right)\right) - 7.506\exp\left(2\left(\xi_2 - \xi_3\right)\right),\\
        \ddot \xi_3 &= 7.522\exp\left(2\left(\xi_2 - \xi_3\right)\right) - 8.406\exp\left(2\left(\xi_3  -\xi_4\right)\right),\\
        \ddot \xi_4 &= 7.900\exp\left(2\left(\xi_3 - \xi_4\right)\right) - 0.00031206\xi_4.
    \end{aligned}
\end{equation}
We observe that the SINDy-predicted dynamical equations for the dark-soliton positions are less accurate
than expected. For instance, the model fails to capture the parabolic-trap contribution for the second and third solitons in the chain. Specifically, the terms $\tfrac{\Omega^{2}}{2}\xi_{2}$ and $\tfrac{\Omega^{2}}{2}\xi_{3}$ should appear on the right-hand sides of $\ddot{\xi}_{2}$ and $\ddot{\xi}_{3}$, respectively. Moreover, the recovered equation,
once again, and in this case more manifestly, violates Newton's third law, as the interaction forces between solitons are not equal and opposite. It is interesting
to observe at an interpretational level that the
resulting regression example appears to trade off
some of the interaction force with the effect
of the parabolic confinement on solitons 2 and 3,
and again to note the effect of the curvature of
the wavefunction profile apparently leading
to (in this case, more substantial) deviations from 
the homogeneous coefficient of $8$ in the case of
the pairwise forces.

For these reasons, we propose a simple, physics-informed alternative to identify a better and more consistent
(also compliant with Newton's third law) set of dynamical equations for the four soliton positions, written in the following matrix form,
\begin{equation}\label{eq: physics-informed dynamics}
    Y = X\beta, 
\end{equation}
where 
\begin{equation}
    \begin{aligned}
        Y &= \left[\ddot\xi_1, \ddot\xi_2, \ddot\xi_3, \ddot\xi_4\right],\\
        X &= \left[\exp\left(2\left(\xi_1-\xi_2\right)\right), \exp\left(2\left(\xi_2 - \xi_3 \right)\right), \exp\left(2\left(\xi_3 - \xi_4\right)\right), \xi_1, \xi_2, \xi_3, \xi_4\right],
    \end{aligned}
\end{equation}
and $\beta$ denotes the following coefficient matrix with the unknown positive entries $a, b, c, d > 0$,
\begin{equation}
    \beta = \begin{bmatrix}
        -a & a & 0 & 0\\
        0 & -b & b & 0\\
        0 & 0 & -c & c\\
        -d & 0 & 0 & 0\\
        0 & -d & 0 & 0\\
        0 & 0 & -d & 0\\
        0 & 0 & 0 & -d
    \end{bmatrix}.
\end{equation}
Essentially, for this choice of $\beta$, we assume that the pairwise interactions between solitons obey Newton’s third law. 
In what follows, we elucidate how Eq.~\eqref{eq: physics-informed dynamics} can be reformulated as a linear regression problem for the unknown variables $[a, b, c, d]$. Specifically, observe that the system \eqref{eq: physics-informed dynamics} can be rewritten as
\begin{equation}\label{eq: transformed system}
    \widetilde{Y} = \widetilde{X}\widetilde{\beta},
\end{equation}
where
\begin{equation}
    \begin{aligned}
        &\widetilde{Y} = \begin{bmatrix}
            \ddot \xi_1\\
            \ddot \xi_2\\
            \ddot \xi_3\\
            \ddot \xi_4
        \end{bmatrix}, \\
      &\widetilde{X} = \begin{bmatrix}
          -\exp\left(2\left(\xi_1-\xi_2\right)\right) &
          \vec{O}_{N\times 1} & \vec{O}_{N\times 1} & -\xi_1\\
          \exp\left(2\left(\xi_1-\xi_2\right)\right) & -\exp\left(2\left(\xi_2-\xi_3\right)\right) & \vec{O}_{N\times 1} & -\xi_2\\
          \vec{O}_{N\times 1} & \exp\left(2\left(\xi_2\xi_3\right)\right) & -\exp\left(2\left(\xi_3-\xi_4\right)\right) & -\xi_3 \\
          \vec{O}_{N\times 1} & \vec{O}_{N\times 1} & \exp\left(2\left(\xi_3-\xi_4\right)\right) & -\xi_4
      \end{bmatrix}, \\
      &\widetilde{\beta} = \begin{bmatrix}
          a, b, c, d
      \end{bmatrix}^T,
    \end{aligned}
\end{equation}
where $\vec{O}_{N\times1}$ denotes the $N$ by $1$ zero column vector.

\begin{figure}[t!]
    \centering
    \includegraphics[width=0.99\linewidth]{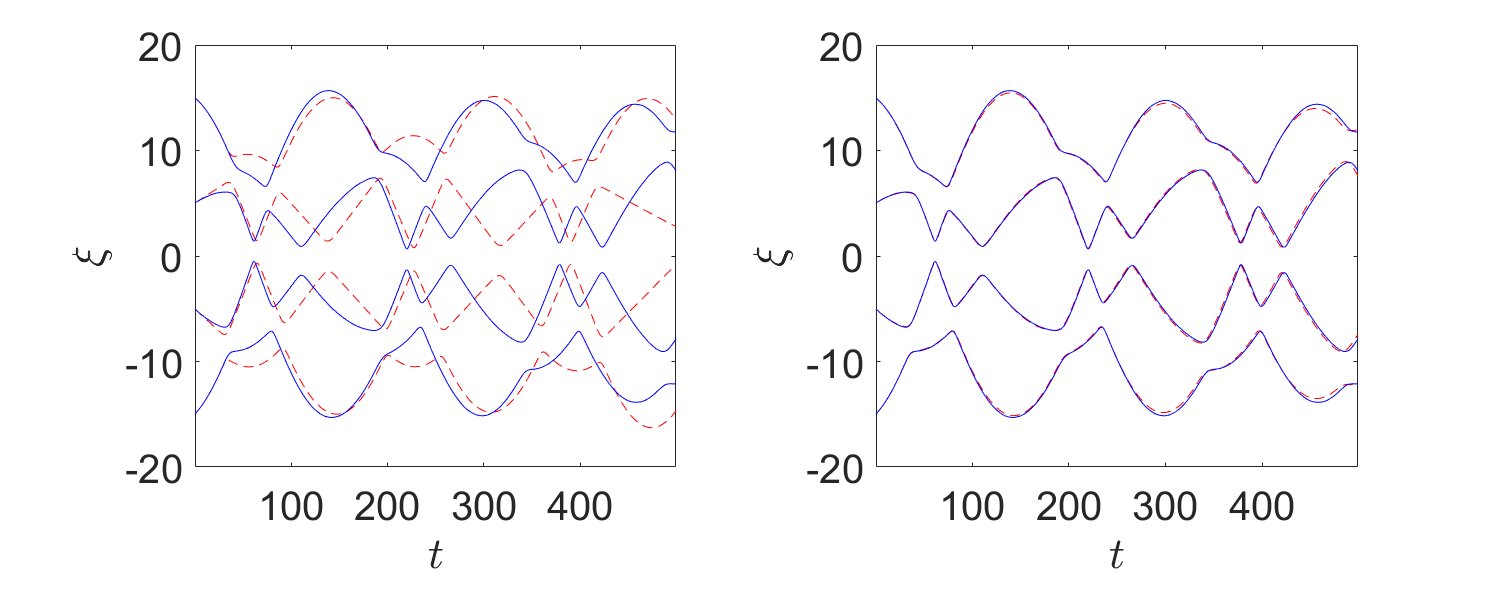}
    \caption{The comparisons of the SINDy predicted dark-soliton positions (based on Eq.~\eqref{eq: dynamics for 4 ds with parabolic trap}) and of the Physics-informed predicted dark-soliton positions (based on Eq.~\eqref{eq: physics-informed learing results}) with that of ground-truth dark-soliton positions sampled from the PDE \eqref{eq: nonlinear schrodinger model}. Notice that the dashed red curves depict the SINDy and Physics-informed predicted trajectories of the dark-soliton positions, while the solid blue curve represents the ground-truth dark-soliton positions sampled from the PDE.}
    \label{fig:SINDy versus physics-informed results}
\end{figure}

We observe that both the left- and right-hand sides of Eq.\eqref{eq: transformed system} are $4N \times 1$ column vectors, where $N$ denotes the length of the time series of $\xi_i$. 
In addition, we note that Eq.~\eqref{eq: transformed system} is a standard linear model, and the solution for $\widetilde{\beta}$ reads, by solving the relevant least square optimization problem, 
\begin{equation}\label{eq: least-square solution}
    \widetilde{\beta} = \left(\widetilde{X}^{T}\widetilde{X}\right)^{-1}\widetilde{X}^{T}\widetilde{Y}.
\end{equation}
Through the numerical implementation of \eqref{eq: least-square solution}, we have (keeping the final numeric results up to the third decimal point) that
\begin{equation}
    \widetilde{\beta} = \begin{bmatrix}
        8.006, & 7.642, & 7.983, & 0.0003176 
    \end{bmatrix}^T.
\end{equation}
Finally, the physics-informed dynamical system of positions of the four dark solitary waves reads
\begin{equation}\label{eq: physics-informed learing results}
    \begin{aligned}
        \ddot \xi_1 &= -8.006\exp\left(2\left(\xi_1 - \xi_2\right)\right) - 0.0003176\xi_1,\\
        \ddot \xi_2 &= 8.006\exp\left(2\left(\xi_1-\xi_2\right)\right) - 7.642\exp\left(2\left(\xi_2 - \xi_3\right)\right) - 0.0003176\xi_2,\\
        \ddot \xi_3 &= 7.642\exp\left(2\left(\xi_2 - \xi_3\right)\right) - 7.983\exp\left(2\left(\xi_3 - \xi_4\right)\right) - 0.0003176\xi_3,\\
        \ddot \xi_4 &= 7.983\exp\left(2\left(\xi_3 - \xi_4\right)\right) - 0.0003176\xi_4.
    \end{aligned}
\end{equation}
Admittedly, Eq.~\eqref{eq: physics-informed learing results} does not perfectly match the theoretically derived reduced-order model for the four-soliton interaction dynamics under a harmonic trap.
However, as per the screening effect of the inhomogeneous
density discussed above (see also~\cite{busch}), this 
level of deviation may be reasonable to expect.
This system already provides a significantly improved recovery of the dynamics compared to its physics-agnostic counterpart, Eq.~\eqref{eq: dynamics for 4 ds with parabolic trap}. Such a claim can also be quantitatively verified by comparing the $L^{\infty}$ loss of the time-series dataset of the soliton positions based on system \eqref{eq: dynamics for 4 ds with parabolic trap} with that of dataset according to system \eqref{eq: physics-informed learing results}. We first notice that the $L^{\infty}$ loss of a given time-series data $\xi = (\xi_{i})_{i=1}^{N}$ is defined as follows,
\begin{equation}\label{eq: L inf loss def}
    L^{\infty}_{\text{loss}}\left(\xi\right) = \max_{1\leq i \leq N}\left(\left|\xi_i - \xi^{\text{GT}}_i\right|\right),
\end{equation}
where $\xi^{\text{GT}}$ refers to the ground-truth data sampled from the original PDE \eqref{eq: nonlinear schrodinger model}. The $L^{\infty}$ loss is $0.478$ for the the system of
Eqs.~\eqref{eq: physics-informed learing results}
(incorporating the relevant physical principles/intuition),
while it is an order of magnitude larger for
the Physics-agnostic system of Eqs.~\eqref{eq: dynamics for 4 ds with parabolic trap}. 
Moreover, Fig.~\eqref{fig:SINDy versus physics-informed results} further demonstrates at the level of trajectories how the Physics-informed prediction \eqref{eq: physics-informed learing results} of the dark-soliton positions performs much better than the ``standard'' SINDy prediction \eqref{eq: dynamics for 4 ds with parabolic trap}.

\subsection{Interaction of bright solitons}\label{subsec: bright solitons results}

We now shift our focus to the interaction dynamics of bright solitons. For conciseness (and having illustrated
some of the issues that arise when going to a larger
number of waves in the dark soliton case), in what follows, we restrict our analysis to the case of up to two interacting bright solitons. Similarly to the scenario of the dark solitons, we also investigate how the presence of a parabolic trapping potential affects the interaction dynamics within the bright-soliton chain.

\subsubsection{One bright soliton with parabolic trap}

We first consider only a single bright soliton 
in the presence of a parabolic trap potential. In this
case, the dynamical equation for
the single bright soliton reads:
\begin{equation}\label{eq: dynamics for BS with parabolic trap}
    \ddot\xi = -\Omega^2\xi.
\end{equation}
The SINDy-identified model for the dynamics of a single bright soliton under the influence of a parabolic trapping potential (with $\Omega=0.025$) is given by:
\begin{equation}
    \ddot\xi = -0.00062491\xi.
\end{equation}
This result is obtained using an extended library that include power of $\xi$ and of $v$ up to order $10$. The fact that SINDy selects only the linear term suggests that it robustly and accurately recovers the true underlying dynamics of the single bright soliton in this 
confined setting.

\subsubsection{Two bright solitons without parabolic trap}

Now, we examine the interaction dynamics of two bright solitons in the case where a parabolic trap is absent. we apply the method to time-series data generated from the following initial position of the two bright solitons:
\begin{equation}\label{eq: two distinct initial positions}
    \begin{aligned}
        \xi_0 = \left[-3, 3\right].
    \end{aligned}
\end{equation}
Notice that the initial positions in system \eqref{eq: two distinct initial positions} mean that the first bright soliton is initially located at the position $x = -3$ while the second bright soliton at $x = 3$.

In these runs, the amplitude is $a = 1$ for both solitons.
The library is: $\Theta = \{x,x^2,\exp\left(x-y\right),\exp\left(-|2L-|x-y||\right)\}$, where $x,y \in \left[\dot\xi_1,\dot\xi_2,\xi_1,\xi_2\right]$. The regular time-series input involves $\dot\xi_1,\dot\xi_2$, while the control variables (since we are once again using
SINDy with control) are $\xi_1,\xi_2$.
The SINDy prediction in this case yields:
\begin{equation}\label{eq: SINDy prediction for [-3,3]}
    \begin{aligned}
        \ddot \xi_1 &= -4.539\exp\left(\xi_1 - \xi_2\right) + 4.327\exp\left(-\left|2L - \left|\xi_1 - \xi_2\right|\right|\right),\\
        \ddot \xi_2 &= 4.539\exp\left(\xi_1 - \xi_2\right) - 4.327\exp\left(-\left|2L - \left|\xi_1 - \xi_2\right|\right|\right),
    \end{aligned}
\end{equation}
Moreover, we have also tried another set of initial positions, which is $\xi_0 = [-2.5, 2.5]$, and the SINDy prediction yields a completely similar dynamics as the system \eqref{eq: SINDy prediction for [-3,3]}.

Here, we find that the SINDy prediction in system \eqref{eq: SINDy prediction for [-3,3]} provides
a reasonable approximation to the theoretically
expected bright soliton interaction; for the latter
the prefactor of Eq.~\eqref{eq: BS reduced dynamical ODE}
would have been equal to $4$, while here we
find it to be closer to $4.5$. Nevertheless,
the interaction accounting for the periodic nature
of the domain is appropriately accounted for and a
reasonably good fit to the dynamics is observed (See the right panel in Fig.~\ref{fig:2 bs with parabolic trap} for such a comparison).

\subsubsection{Two bright solitons with parabolic trap}

We also investigate the interaction dynamics of two bright solitons in the presence of a parabolic trapping potential.

\begin{figure}[t!]
    \centering
    \includegraphics[width=0.45\linewidth]{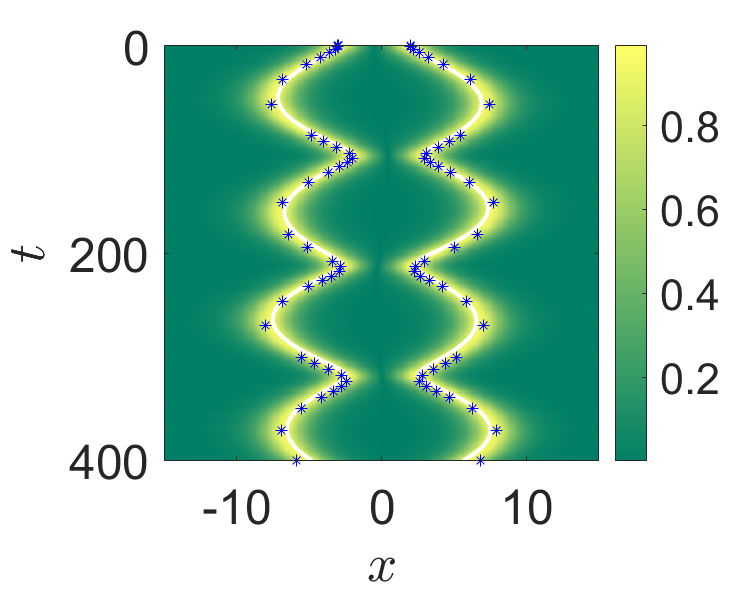}
    \includegraphics[width=0.45\linewidth]{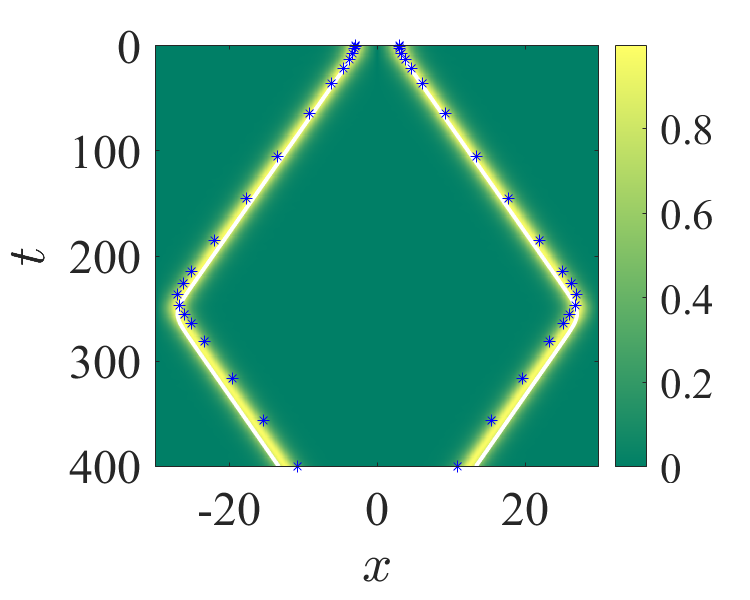}
    \caption{The spatio-temporal plot of the bright-soliton interaction in the presence of (Left panel) and without (Right panel) a parabolic trap potential. Notice that the solid white curves and the blue asterisks depict the theoretical prediction of the positions of the two bright solitons based on system \eqref{eq: dynamics for BS with parabolic trap} and the SINDy prediction on the soliton positions based on \eqref{eq: SINDy dynamics for 2BS with trap} and \eqref{eq: SINDy prediction for [-3,3]}.
    The underlying contour map stems from the simulation of the 
    original
    NLS PDE with the parabolic confinement.}
    \label{fig:2 bs with parabolic trap}
\end{figure}

The SINDy-identified model for this scenario is given by:
\begin{equation}\label{eq: SINDy dynamics for 2BS with trap}
    \begin{aligned}
        \ddot\xi_1 &= -4.607\exp\left(\xi_1-\xi_2\right) - 0.000619\xi_1,\\
        \ddot\xi_2 &= 4.602\exp\left(\xi_1-\xi_2\right) - 0.000615\xi_2.
    \end{aligned}
\end{equation}
{As one can see in Figure~\ref{fig:2 bs with parabolic trap}, when there is a harmonic trap, the SINDy prediction is 
reasonably accurate, matching well with the associated analytical approximated dynamics described in Eq.~\eqref{eq: OOP bright solitons interaction with trap} and importantly
with the underlying PDE contour map also
depicted in the Figure.}

\subsubsection{Multiple degrees of freedom: A case example}\label{subsubsec: case example}

Up to now, all previous numerical experiments have involved systems with only two degrees of freedom. Namely, the dynamics systems only consider the positions $\xi$ and the velocities $v$ of the solitons. To explore the capabilities of SINDy on higher-dimensional systems and also to illustrate
some of the methodological limitations of the present
approach, we now turn our attention to the full dynamical model with four degrees of freedom in the case of
two bright solitons, 
as described by Eq.~\eqref{eq: BS dynamical ODEs BSs}.

Similarly to what was discussed before, we generate all the necessary time-series data for all four relevant parameters of the bright solitons, namely the amplitude, 
position, velocity and phase thereof. Subsequently, we examine SINDy’s performance in this more complex setting. Due to the increased dimensionality, we restrict our investigation to the case of two interacting bright solitons, assuming they are out-of-phase, i.e., we have that $\sigma_{i,j} = 1$. Admittedly, this case is somewhat
simpler to tackle than the one of the in-phase
attractive interaction, especially so in the context
of non-integrable models (where the solitons may 
exchange mass etc.). While these complications
are worthwhile to consider in future work, we will
steer clear of them here.

In the present case, the SINDy prediction for this four degrees-of-freedom dynamical system reads
\begin{equation}\label{eq: SINDy prediction on 4-degrees of freedom system}
    \begin{aligned}
        &\dot a_1 = 0.841(-4a_1^{2}S_{12}),\quad \dot a_2 = 1.040(4a_2^{2}S_{21}),\\
        &\dot v_1 = 0.733 (6a_2C_{21}), \quad \dot v_2 = -0.769(6a_2C_{21}),\\
        &\dot\xi_1 = 1.000v_1,\quad \dot\xi_2 = 1.000v_2,\\
        &\dot\phi_1 = 0.999\left(\frac{a_1^{2}+v_1^{2}}{2}\right), \quad \dot\phi_2 = 0.999\left(\frac{a_2^{2}+v_2^{2}}{2}\right).
    \end{aligned}
\end{equation}
Notice that in this case the library is $\Theta = \{x\}$ where 
\begin{equation}
x \in \left[-4a_1^2S_{12}, 4a_2^2S_{21}, 4a_1^2C_{12}, -4a_2^2C_{21}, v_1, -2S_{12}, v_2, -2S_{21}, \frac{1}{2}\left(a_1^2+v_1^2\right), -2v_1S_{12}, 6a_1C_{12},
\frac{1}{2}\left(a_2^2+v_2^2\right), -2v_2S_{21}, 6a_2C_{21} \right].
\end{equation}
All these terms are used as control variables in SINDy. Finally, for this multi-degree case example, the optimizer we used was the usual greedy algorithm FROLS.

\begin{figure}[t!]
    \centering
    \includegraphics[width=0.88\linewidth]{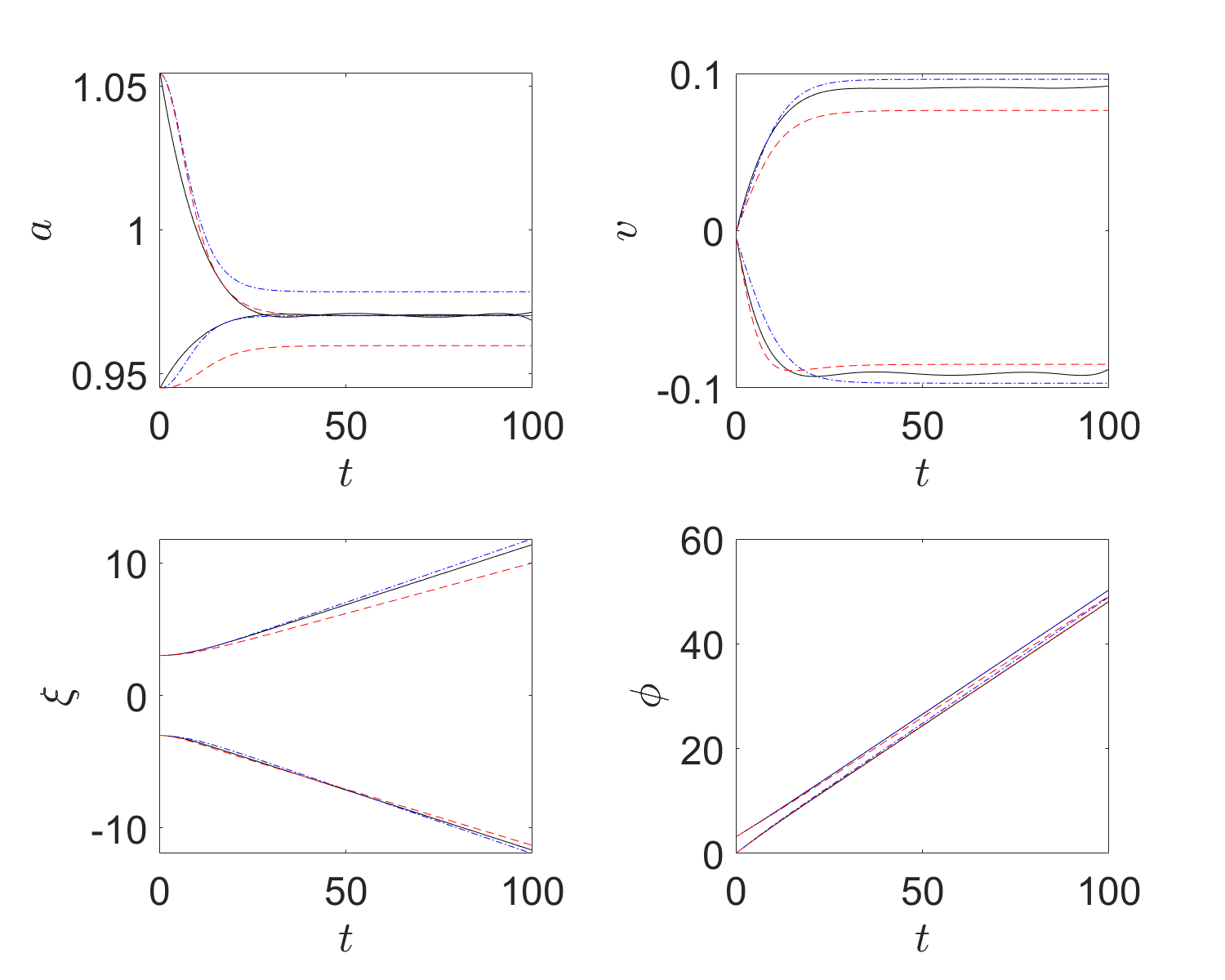}
    \caption{The comparison of the four soliton parameters in the four degrees of freedom system: we notice that in each panel above, the solid black, dashed red, and dashed-dotted blue curves depict the time-series data generated from the PDE simulation, the theoretical prediction based on system \eqref{eq: BS dynamical ODEs BSs}, and SINDy prediction based on the system \eqref{eq: SINDy prediction on 4-degrees of freedom system} for each of the two solitons. }
    \label{fig: comparison of the 4 parameters of the bright solitons}
\end{figure}

{It is worthwhile to note that unfortunately, in this more complex, multi-degree case example, SINDy fails to yield a prediction which closely aligns with the corresponding approximated dynamics listed in system \eqref{eq: BS dynamical ODEs BSs}. However, based on its prediction, as 
depicted in Figure \ref{fig: comparison of the 4 parameters of the bright solitons}, SINDY's predicted dynamics (in dashed-dotted blue) is still capable to qualitatively capture the evolution dynamics of the four distinct characteristics of the solitons (in solid black, as obtained from the 
PDE simulation). Moreover, it is also worthwhile to notice that in this particular example, the theoretical prediction based on system \eqref{eq: BS dynamical ODEs BSs} also deviates from the actual dynamics. 
Indeed, one can argue that the SINDy results are comparably
good, if not better than the ones obtained variationally.}

On the other hand, there are also certainly 
``shortcomings'' in the obtained expressions which may
at least partially inhibit their generalizability.
For instance, once again, one would expect the impact
of the presence of the 2nd soliton on the 1st to be
the same as that of the first on the 2nd. In that sense,
the equations on the first two lines do not reflect
that reciprocity. The ones on the 3rd and 4th line
seem to accurately reflect the dominant dynamics,
but again (e.g., in the 4th line) there exists a
potential shortcoming in that the greedy optimizer
only identifies the dominant contribution to the
phase dynamics, neglecting the exponentially weaker
2nd and 3rd terms in the 4th line of Eq.~\eqref{eq: BS dynamical ODEs BSs}. While this does not appear to 
contribute significantly here, there may exist
settings where such a term may play a more significant
role. On balance, here we find that there is more
room for potential improvement. The incorporation of
physical attributes of the dynamics (such as the potential
force reciprocity etc.) may enable further refinement
of the dynamics identified. Nevertheless, we trust
that this array of progressively more complex examples
will be convincing to the reader of the potential
value of utilizing such methods as a complement to
existing techniques towards identifying the 
projections on the solitonic manifold (potentially even
prior to seeking to analytically derive ones such).

\section{Conclusions and future directions}\label{sec: Conclusions}

In this work, we have revisited the time-honored
problem of deriving effective equations 
for fundamental wave-like solutions of dispersive
nonlinear partial differential equations, such as, notably,
the bright and dark soliton solutions of the NLS equation.
This so-called projection on the solitonic manifold 
and corresponding 
study of the resulting single-soliton and interaction dynamics of multiple solitons has been approached
from a complementary data-driven perspective. 
While it is not inconceivable that such a data-driven
approach may yield results more accurate in special cases
(yet, perhaps, often not as generalizable) as the 
traditional variational and asymptotic paper-and-pencil
methods, its role here is intended to be complementing and
perhaps guiding those, rather than replacing them.
More specifically, in our considerations, 
we focused on identifying systems of ODEs that approximately govern the evolution of key characteristics of the solitonic
waveforms. To this end, we employed the SINDy algorithm and its extension, SINDy with control, to learn the governing equations directly from time-series data. This was typically
carried out for the position and velocity pair of
conjugate variables, although some examples were also
given of bright soliton cases where more variables
(including amplitude and phase) were also involved.

Broadly speaking,
our numerical experiments demonstrated that SINDy is capable of accurately recovering the theoretically derived reduced-order models for a variety of soliton interaction scenarios. However, we also observed limitations: in particular, the
method could benefit from incorporation of physical 
principles such as the reciprocity of interactions 
between two solitons.
Furthermore, the method showed reduced accuracy when applied to higher-dimensional systems, as illustrated in the
multi-parameter, bright soliton case study presented in Subsubsection~\ref{subsubsec: case example}. This degradation in performance is likely attributed to the increased complexity of the underlying model (and the 
competition therein of multiple effects at
different orders). Indeed, it is important to recognize the inherent challenges of applying sparse regression techniques in high-dimensional settings. Nevertheless, these limitations 
pose challenges that will be relevant to consider 
in the future, possibly aided by more refined
data-driven/machine learning tools.

In terms of the future directions, one possible 
avenue is to try to utilize alternative machine learning approaches such as the Neural Ordinary Differential Equations (Neural ODEs)~\cite{chen2019neuralordinarydifferentialequations} for dynamical equation identification when the reduced-order model has a relatively large dimensionality (e.g. see \ref{subsubsec: case example}). Another direction is to extend the study of soliton interaction dynamics to some other mathematical physics models, such as the KdV equation mentioned in the introduction \ref{sec: introduction}
and observe the outcomes therein. 
Indeed, upon amassing sufficient experience with known
models (such as NLS, KdV, sine-Gordon etc.), one hopes to
utilize this toolbox also in problems (possibly also 
higher-dimensional ones etc.) where such equations may not
be analytically known.
Additionally, while the current study focuses on interactions involving  four or fewer solitons (Section~\ref{sec: numerical results}), investigating the dynamics of larger soliton ensembles would provide a more rigorous test of the robustness of the techniques utilized herein. Although the underlying dynamics described by Eqs.~\eqref{eq: dark solitary wave dynamics ODE} and \eqref{eq: BS reduced dynamical ODE} are expected to hold, such extended studies could further validate the consistency and scalability of data-driven model discovery. A number of  these avenues is currently in progress and relevant results 
will be reported in future publications.

\section*{Acknowledgments}
This material is based upon work supported by the U.S. National Science Foundation under the award PHY-2110030, PHY-2408988 and DMS-2204702 (P.G.K.) and DMS-2502900 (WZ) and by the Air Force Office of Scientific Research (AFOSR) under Grant No. FA9550-25-1-0079 (WZ).
This research was partly conducted while P.G.K. was 
visiting the Okinawa Institute of Science and
Technology (OIST) through the Theoretical Sciences Visiting Program (TSVP). 
This work was also 
supported by a grant from the Simons Foundation
[SFI-MPS-SFM-00011048, P.G.K].

\bibliographystyle{unsrt}

\bibliography{main}

\end{document}